# WEB SERVICES DEPENDENCY NETWORKS ANALYSIS

Chantal Cherifi[1,2], Vincent Labatut[1]
[1]*Computer Science Department, Galatasaray University, Ortaköy, Istanbul, Turkey*
*chantalbonner@gmail.com, vlabatut@gsu.edu.tr*

Jean-François Santucci[2]
[2]*SPE Laboratory, UMR CNRS, University of Corsica, France*
*santucci@univ-corse.fr*

Abstract: Along with a continuously growing number of publicly available Web services (WS), we are witnessing a rapid development in semantic-related web technologies, which lead to the apparition of semantically described WS. In this work, we perform a comparative analysis of the syntactic and semantic approaches used to describe WS, from a complex network perspective. First, we extract syntactic and semantic WS dependency networks from a collection of publicly available WS descriptions. Then, we take advantage of tools from the complex network field to analyze them and determine their topological properties. We show WS dependency networks exhibit some of the typical characteristics observed in real-world networks, such as small world and scale free properties, as well as community structure. By comparing syntactic and semantic networks through their topological properties, we show the introduction of semantics in WS description allows modeling more accurately the dependencies between parameters, which in turn could lead to improved composition mining methods.

## 1 INTRODUCTION

A Web Service (WS) is a self-describing, self-contained, modular application accessible over the web. It can be published and discovered in a WS registry and invoked for remote use. Its interface is exposed throughout a so-called WS description, which lists the implemented operations. Currently, production WS use syntactic descriptions expressed with the most prevalent description language, WSDL, a W3C (World Wide Web Consortium) standard. Along with textual information and some low level access directives, descriptions basically contain the names of the operations and their parameters names and data types. WS were initially designed to interact with each other, in order to provide a composition of WS able to offer higher level functionalities (Benatallah et al., 2003). The keyword-based techniques used in current production discovery are not suitable as they often lead to false positives and false negatives (Pilioura et al., 2003). A false positive takes the form of an irrelevant service which includes the searched keywords in its description. On the contrary, different descriptions can contain syntactically different but semantically equivalent words, leading to false negatives. Furthermore, keyword-based techniques do not allow to perform any form of inference nor flexible match (Sycara et al., 2003). The key underlying problem is that keywords are a poor way to capture the semantics of a search or an advertisement. More advanced research (non-production) approaches rely on comparing structured data such as parameters types and names, or analyzing unstructured textual comments (Stroulia et al., 2003; Wu et al., 2005; Ma et al., 2008). This is generally not enough to distinguish WS in terms of functionality, and consequently makes it difficult, or even impossible, to use these methods to automate WS composition. Indeed, syntactically discovered WS must be manually validated to ensure they implement the desired behavior, leading to static, a priori compositions.

To solve this limitation, a different mechanism is needed, one that entails retrieving WS on the basis of the functionalities they provide. The research community followed the current semantic Web trend by introducing semantics in WS descriptions, in order to enrich them. Several initiatives for new semantic description languages exist among which we can distinguish purely semantic descriptions with OWL-S (also a W3C standard), from annotated WSDL descriptions with WSDL-S and SAWSDL. Although those languages allow associating ontological concepts with various elements of the description, the research community has been focusing only on those qualifying the operations inputs and outputs. But retrieving semantic information is far more costly than collecting syntactic descriptions, even when considering only parameters. The latter can be performed quickly and completely automatically, whereas the former is a long task, requiring human intervention to label each parameter with the proper concept. Annotation tools exist to help, but they are clearly not mature yet, and often defined for specific collections or languages (Hess et al., 2004; Gomadam et al., 2005). Maybe for these reasons, no semantic annotation language emerged as an industry standard, although they appeared more than five years ago now: all production WS still rely on WSDL. Even at a research level, no publicly available significant collection of semantically annotated WS exists, making it very difficult to test new algorithms.

This situation leads to one question: is describing WS semantically worth the cost? To our knowledge, no one did ever compare the information underlying syntactic and semantic WS descriptions. In this work, we try to tackle this problem from the perspective of parameters dependency, through the use of complex networks. We model parameters spaces by building so-called dependency networks, based on syntactic and semantic descriptions of a single WS collection. We make the assumption the information conveyed by the two different kinds of descriptions appears in the corresponding dependency networks. We then compare the syntactic and semantic descriptive approaches through the networks topological properties. Our main contributions are an extended investigation of the parameters networks topology and the comparison of syntactic and semantic networks. The rest of the paper is organized as follow. In section 2, we present complex networks and their main topological properties. Section 3 introduces dependency networks and explains how they can be extracted from WS descriptions. Section 4 is dedicated to the presentation and discussion of our experimental results, i.e. the extracted networks, their topological properties and how they compare. Finally, in section 5, we emphasize the original points of our work, discuss its limitations and their possible solutions, and explain how it can be carried on.

## 2 COMPLEX NETWORKS

Complex networks are a specific class of graphs, characterized by a huge number of nodes and non trivial topological properties. They are used in many different fields to model real-world systems (Costa et al., 2008), and have been intensively studied both theoretically and practically (Newman, 2003). Because of their complexity, specific tools are necessary to analyze and compare them. This is usually performed through the comparison of several well-known properties, supposed to summarize the essential of the network structure.

### 2.1 Distance-Based Measures

The *distance* between two nodes is defined as the number of links in the shortest directed path connecting them. At the level of the whole network, this allows to process the *average distance* and the diameter. The former corresponds to the mean distance over all pairs of nodes (Newman, 2003). This notion is related to the *small-world* property, observed when this distance is relatively small. The classic procedure to assess this property consists in comparing the average distance measured in some network of interest to the one estimated for an Erdős–Rényi (ER) network (Erdos et al., 1959) containing the same numbers of nodes and links, since this random generative model is known to produce networks exhibiting the small-world property (Newman, 2003). In terms of dynamic processes, the existence of shortcuts between nodes can be interpreted as propagation efficiency (Watts et al., 1998). Most real-world networks have the small-world property. The *diameter* is the greatest distance over all pairs of nodes in the network. In real-world networks, a small diameter is synonymous to rapid information propagation (Cherifi, 2005).

### 2.2 Transitivity

A network *transitivity* (also called clustering) corresponds to its density of triangles, where a triangle is a structure of three completely connected nodes. It is measured by a so-called transitivity coefficient, which is the ratio of existing triangles to possible triangles in the considered network (Watts and Strogatz, 1998). The higher this coefficient, the more probable it is to observe a link between two nodes which are both connected to a third one. A real-world network is supposed to have a higher transitivity than the corresponding ER network by an order of magnitude corresponding to their number of nodes, meaning their nodes tend to form densely connected groups.

## 2.3 Degree-Based Measures

The *degree* of a node corresponds to the number of links attached to it. In a directed network, one can distinguish in and out degrees, i.e. the numbers of incoming and outgoing links, respectively. Nodes with a high in (resp. out) degree are called authorities (resp. hubs). The most basic degree-based measure is certainly the *average degree* over the whole network. When comparing networks containing the same number of nodes, it is related to their link density. The degree distribution of a network is particularly revealing of its structure. Most real-world networks have a power law degree distribution (Albert et al., 1999; Newman, 2003; Boccaletti et al., 2006), resulting in the so-called *scale-free* property. In other terms, real-world networks contain a very few nodes with extremely high degree, and a large number of nodes with very small degree.

The *degree correlation* of a network constitutes another interesting property. The question is to know how a node degree is related to its neighbors'. Real networks usually show a significantly different from zero degree correlation. If it is positive, the network is said to have assortatively mixed degrees whereas if it is negative, it is disassortatively mixed (Newman, 2003). According to Newman, social networks tend to be assortatively mixed, while other kinds of networks are generally disassortatively mixed.

## 2.4 Component Organization

A *component* is a maximal connected sub-graph, i.e. a set of interconnected nodes, all disconnected from the rest of the network. The component distribution and, more specifically, the size of the largest component are important network properties. Indeed, depending on the applicative context, the fact the network is split in several separated parts with various sizes can be considered as an indirect representation of the modeled system effectiveness. For example, in a communication network like the Internet, the size of the largest component represents the largest fraction of the network within which communication is possible and hence it reflects the effectiveness of the network at doing its job (Newman, 2003). Most real-world networks have a so-called *giant component*, whose size is overwhelming greater than the other components.

## 2.5 Community Structure

A *community* is defined as a subset of nodes densely interconnected relatively to the rest of the network. Unlike components, communities are not necessarily disconnected from each other (and generally, they are significantly connected). Many real-world networks have a community structure (Newman, 2003). Specific community detection algorithms must be used to identify them, leading to a partition of the overall nodes set. Almost all of them are dedicated to undirected networks, and only a very few recent ones can use the information conveyed by directed links. In this work, we chose to use a well tested program, and therefore focused on undirected links. We selected the Walktrap algorithm which exhibits good performances according to recent benchmarks (Orman et al., 2009).

To assess the quality of a network partition, the standard measure is Newman's *modularity* (Newman et al., 2004), whose value also depends on the considered network structure. Consequently, its theoretical maximal value of 1 (perfect community structure and partition) is rarely reached, and in practice values between 0.3 and 0.7 are considered high (Newman, 2006). A value of 0 represents a random partition or the absence of community structure.

## 3 DEPENDENCY NETWORKS

## 3.1 Network Definition

We define a dependency network as a directed graph whose nodes correspond to depending objects and links indicate the head nodes depends on the tail nodes. They can be considered as complex networks, and a few authors used similar approaches to model collections of WS based on syntactic (Kil et al., 2006; Oh, 2006) or on semantic (Hashemian et al.,

2005) descriptions. In the resulting parameters networks, each node corresponds to a parameter, and the links between them represent operation-related dependences. In this work, our goal is to compare the two types of WS descriptions; hence we used both syntactic and semantic descriptions.

As stated before, a WS interface is defined under the form of a set of operations. An operation $op_i$ represents a specific functionality, described independently from its implementation for interoperability purposes. Besides its functionality, it is characterized by two sets of input and output parameters, noted $I_i$ and $O_i$, respectively. In a syntactic description, each parameter has a name and a type. This type is also defined independently from any implementation, again for interoperability reasons. Most of the time, the XML schema definition language (XSD) is used. In a semantic description, name and type are also generally specified, and an additional ontological concept is associated to the parameter, in order to give it a precise meaning. The most popular language used to describe these concepts is OWL, the Web Ontology Language designed by the W3C Web Ontology Working Group.

In the context of dependency networks, each operation $op_i$ is formally defined as a triplet $(I_i, O_i, K_i)$ (Hashemian and Mavaddat, 2005), where $K_i$ denotes the set of dependencies defined by the operation. We consider each output parameter depends on each input parameter. The left side of Figure 1 represents three operations $op_1$, $op_2$ and $op_3$, with their respective inputs and outputs under the form of 9 parameters named with letters ranging from $a$ to $i$. As an example, consider operation $op_2$: it is defined as $op_2 = (I_2, O_2, K_2)$ where: $K_2 = \{(c,e),(c,f),(d,e),(d,f)\}$, $I_2 = \{c,d\}$ and $O_2 = \{e,f\}$ (i.e. $e$ and $f$ are both dependent on $c$ and $d$). When considering not only a single operation, but a whole collection, one can say a parameter $p_2$ depends on another parameter $p_1$ iff an operation $op_i$ exists such as $p_1 \in I_i$ and $p_2 \in O_i$.

Parameters dependency network capture this three-part information: their nodes represent the parameters ($I_i$ and $O_i$) and their links stand for the dependencies ($K_i$). To build such a network, we first create one node for each parameter present in the whole collection. Then, links are created by considering each operation separately: a link is added between each one of its input parameters and each one of its output parameters. The right side of Figure 1 represents the dependency network corresponding to the three operations described on the left side. For example, for each input parameter $\{a,b\}$ belonging to $op_1$, there exists a link directed to each one of its output parameters $\{c,d,e\}$. In this network, the presence of a link from a node $p_1$ towards another node $p_2$ indicates at least one operation uses the parameter corresponding to $p_1$ as an input, and the parameter corresponding to $p_2$ as an output. This can also be interpreted in terms of *production*: we say such a link means one or several operations allow producing $p_2$ provided $p_1$ is already available.

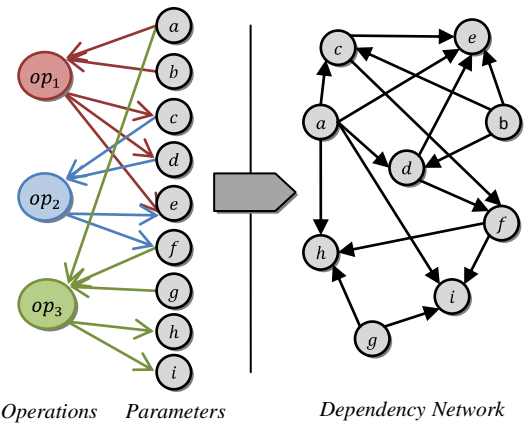

*Operations    Parameters         Dependency Network*

Figure 1: Example of dependency network extraction.

Connectivity in a dependency network is caused by the fact one parameter may be used by several operations, either as an input or an output. For example, parameters $\{a,c,d,e\}$ appear more than once in the collection, either as inputs or outputs for several operations, but only one node stands for each of them in the resulting dependency network. The parameters described in the collection of WS descriptions (left side in Figure 1) are called parameter *instances*. Those represented by nodes in the dependency network (right side) are called parameter *archetypes*. One parameter archetype represents a group of parameter instances supposed to convey the same information. Consequently, deciding if two instances correspond to the same archetype is a central task in extracting a dependency network. This depends on the nature of the considered parameters (syntactic vs. semantic description) and on how the notion of similarity is defined. These factors are implemented under the form of a so-called matching function.

## 3.2 Matching Function

A *matching function* $f$ takes two parameters $p_1$ and $p_2$, and determines their level of similarity (Shvaiko et al., 2005), generally under the form of a value in

[0,1]. This function can be either symmetrical ($f(p_1, p_2) = f(p_2, p_1)$) or asymmetrical, and its output can be either binary or real. When comparing two parameters, a real output allows representing various levels of similarity, which is a good thing because it conveys a more subtle information than a raw binary value. But it also results in a more complex processing during network generation. So we selected only binary matching functions in order to avoid this situation.

Because of the different nature of the concerned information, we used different matching functions to compare syntactically and semantically described parameters, resulting in so-called syntactic and semantic dependency networks, respectively. For syntactic descriptions, we compare parameters names: two parameters are said to be similar if their names are the exact same strings. The semantic matching is performed against the concepts associated to the parameters. It is based on the classic *exact* operator used in previous WS-related works to compare ontological concepts (Paolucci et al., 2002). It considers two parameters to be identical iff their associated concepts match perfectly. Note both matching functions are symmetrical. Our goal is to compare syntactic and semantic descriptions, not matching functions, so we opted for standard and simple tools. In summary, we can extract two distinct networks: *syntactic equal* and *semantic exact*, noted $N^{Eq}$ and $N^{Ex}$, respectively.

## 4 RESULTS AND DISCUSSION

### 4.1 Data

We extracted interaction networks from the SAWSDL-TC1 collection of WS descriptions (Klusch et al., 2008; "SAWSDL-TC", 2008). This test collection provides 894 semantic descriptions written in SAWSDL, and distributed over 7 thematic domains (education, medical care, food, travel, communication, economy and weapon). It originates in the OWLS-TC2.2 collection, which contains real-world services descriptions retrieved from public IBM UDDI registries, and semi-automatically transformed from WSDL to OWL-S. This collection was subsequently resampled to increase its size, and converted to SAWSDL. An SAWSDL file describes a service both syntactically and semantically. This allowed us to extract our syntactic and semantic networks from the same collection, and to compare them consistently.

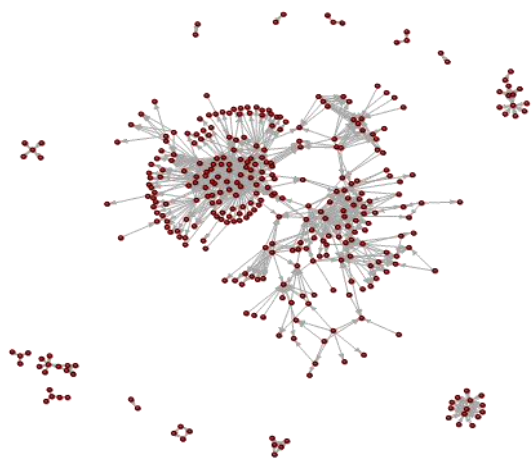

Figure 2: Trimmed exact semantic network $N^{Ex}$. The giant component is located in the middle.

### 4.2 Networks Structure

The whole SAWSDL-TC1 test collection contains 2136 parameters instances, represented by 385 nodes in $N^{Eq}$ and 357 nodes in $N^{Ex}$. The proportion of isolated nodes, 4.67% and 4.2% in $N^{Eq}$ and $N^{Ex}$ respectively, is quite small in both networks, and even a little smaller in the semantic one. Both networks exhibit a giant component. In $N^{Eq}$ it contains 73% of the remaining nodes and 86% of the remaining links. The others 17 components are much smaller ranging from 2 to 30 nodes. $N^{Ex}$ is separated in 16 distinct components, the giant one containing 78% of the nodes and 88% of the links (see Figure 2).

The fact that many parameters instances appear many times in the collection, like for example, `_PRICE` (130 occurrences) or `_AUTHOR` (74 occurrences), is the reason why there is a significant difference between the number of instances and the number of parameter archetypes (i.e. nodes in the networks). Moreover, we used different matching functions to build the syntactic and semantic networks, so the resulting archetypes are different (i.e. they do not correspond to the same sets of instances), which explains the difference in the number of nodes between $N^{Eq}$ and $N^{Ex}$. The number of nodes is smaller for the later, which indicates semantic matching allows associating more instances.

This highlights the presence of *false negatives* (FN) in the syntactic network. FN are instances associated to different archetypes in the dependency network, whereas they are actually conveying the same information, and should therefore be represented by the same archetype. These FN usually are syntactically different (different names) but are associated to the same ontological concept (same meaning). For example parameter instances `_AUTHOR`, `_AUTHOR1` and `_AUTHOR2` are represented by three distinct nodes in the syntactic network, whereas they are associated to a unique node in the semantic network, as they all are associated to the same `#author` concept. The semantic matching also allows eliminating some *false positives* (FP). FP correspond to instances represented by the same archetype whereas they do not represent the same information. For example, many instances are simply called `PARAMETER` but are associated to very different concepts. The syntactic matching will improperly associate them to a common archetype, whereas the semantic matching will not.

Globally, the semantic matching results in less isolated nodes and small components, and a larger giant component, both in terms of nodes and links. The fact distinct components exist reflects the decomposition of the collection into several non-interacting groups of parameters. The presence of the giant component is a good property. It means the number of dependencies in which several operations are implied is high, allowing a large proportion of parameters to interact. In the rest of this section, we focus on the giant components properties, discarding isolated nodes and smaller components. Table 1 lists which properties we discuss hereafter. For average distance and diameter, values are given for both directed and non-directed networks.

### 4.3 Distance-Based Measures

Both syntactic and semantic networks exhibit small average distances: 2.75 and 1.97, respectively. By comparison, this distance is approximately 6 in ER random networks of comparable size, which means the dependency networks possess the small world property. In other terms, many shortcuts exist in the networks, indicating one can find dependency paths using a relatively small number of parameters. This can be interpreted in terms of WS composition, meaning one can produce some parameters of interest using a relatively small number of operations. According to the results, we can say semantic matching generates a more distinct small-world property.

The component diameter is a good indicator of the largest dependency path, which is 7 for $N^{Eq}$ and 5 for $N^{Ex}$. Observing this significant difference between the syntactic and the semantic networks allows us to say producing parameters is more efficient in terms of number of required operations. The fewer operations there are, the shorter the production time and the smaller the chance to meet unavailable operations on the path. Another point is the directed nature of the network, which leads to sensibly different results: from 3 to 4 and from 2 to 4 for the respective average distances of $N^{Eq}$ and $N^{Ex}$, and from 7 to 10 and 5 to 9 for their respective diameters.

Table 1: Properties of the giant components.

| Property | $N^{Eq}$ | $N^{Ex}$ |
|---|---|---|
| Nodes | 269 | 268 |
| Links | 633 | 621 |
| Average distance (directed / undirected) | 2.75 / 4.01 | 1.97 / 4.18 |
| Diameter (directed / undirected) | 7 / 10 | 5 / 9 |
| Transitivity | 0.039 | 0.031 |
| Degree correlation | −0.21 | −0.22 |
| Average degree (in / out / all) | 2.3 / 2.3 / 4.7 | 2.3 / 2.3 / 4.6 |
| p-value (in / out / all) | 0.42 / 0.02 / 0.81 | 0.57 / 0.21 / 0.84 |
| Communities | 16 | 16 |
| Modularity | 0.62 | 0.62 |

### 4.4 Transitivity

Unlike most real-world networks, the measured transitivity is relatively low for both syntactic and semantic networks, with values around 0.03. By comparison, this transitivity coefficient is approximately 0.02 in both ER random networks of comparable size, known to have very low transitivity. As shown in Figure 2, parameters are organized very hierarchically, in the form of trees rather than triangles, which explains these low values. This structure favors the apparition of hubs and authorities, the former corresponding to parameters used as an input by many operations and the latter to parameters being outputs for many operations. They play a central role in the parameters production process, and their absence can be critical. If a parameter is a hub, the production of many others depends on its presence. If it becomes unavailable, all these parameters cannot be produced anymore. If a parameter is an authority, its production depends on many others; and there are many operations able to produce it. For example, `_COUNTRY` and `_PRICE` are such remarkable parameters used or produced by several operations.

## 4.5 Degree-Based Measures

An empirical analysis of the network shows few parameters have a huge number of links while the majority has only a few, which is characteristic of a power law degree distribution. To confirm this observation, we used the method proposed in (Clauset et al., 2009) to fit our data. We obtained high p-values for the global degree distribution (all) as showed in Table 1. Hence, the null hypothesis cannot be rejected, allowing us to suppose this distribution follows a power law (see Figure 1).

The p-values are lower for the in and out degrees distributions. They still do not allow rejecting the power law distribution, except for the syntactic out degree if we take a threshold of 0.05. Indeed, one can observe the existence of few hubs and authorities.

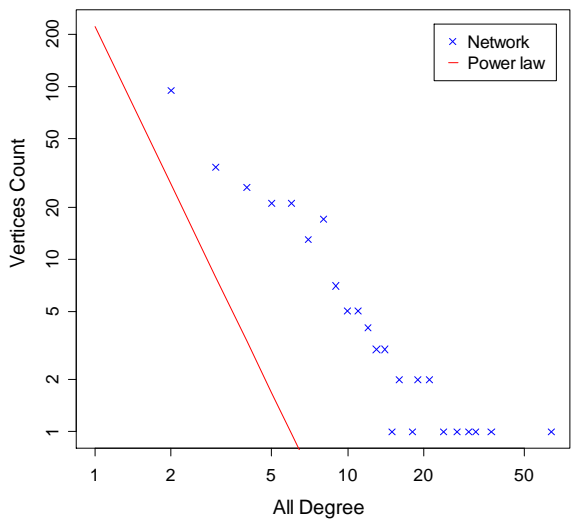

Figure 3: Degree distribution for $N^{Ex}$, on a log-log scale.

These results are in accordance with average degree and degree correlation values. In Table 1, negative values of degree correlation indices indicate that nodes are significantly disassortatively mixed. Strongly connected nodes, as hubs and authorities are preferentially linked with lightly connected ones. While the value for the maximum degree is 63 in both $N^{Eq}$ and $N^{Ex}$, the average degrees values are relatively low, due to the presence of a high number of poorly connected nodes.

## 4.6 Community Structure

The Walktrap algorithm detected communities with a good modularity (0.62) for equal and exact networks (values greater than 0.3 are considered high (Newman, 2006). This community structure seems to reflect the collection domains, i.e. there is a partial correspondence between the groups of parameters retrieved from the network structure and those belonging thematically to specific domains. Indeed, it makes sense to observe denser relationships between parameters belonging to the same application field, because it is likely the related operations were designed to interact with each other. Parameters which are responsible to bridges between communities may be remarkable nodes such as hubs and authorities. For example `_COUNTRY` and `_PRICE` are transverse to the collection, i.e. they are produced or used by many operations across several domains. We did not notice any significant difference between the two considered networks while observing community structure.

## 5 CONCLUSIONS

In this paper, we compared the information conveyed by syntactic and semantic WS descriptions, through the use of complex networks. For this purpose, we extracted a syntactic and a semantic dependency networks from one collection of descriptions, using two different matching functions. We processed, discussed and compared their topological properties. Both networks exhibit some properties observed in most real-world complex networks: small average distance, power law degree distribution, presence of a giant component and community structure. However, unlike most real-world network, our dependency networks are not highly transitive.

When comparing syntactic and semantic networks, we observed a greater proportion of nodes and links are included in the semantic giant component. Consequently, the number and the size of small components decrease as well as the number of isolated nodes. The semantic giant component interconnection structure leads to smaller average distance and diameter. This means one needs to chain fewer operations to produce a given parameter. We can conclude the introduction of semantics in WS description allows a more accurate representation of their dependency relations.

Dependency network-based representations of WS collections were used before in various contexts. (Hashemian and Mavaddat, 2005) used them for composition mining, i.e. finding the best composition relatively to some criteria of interest. Oh *et al*. elaborate a benchmark dedicated to WS discovery and composition (Oh et al., 2009) and developed a WS composition algorithm (Oh et al., 2008). The two latter works are based on a study of networks topological properties (Kil, Oh et al., 2006). However, this study focused only on syntactic descriptions, and neither the directed nature of the links nor the community structure are considered, Yet, this is of great importance in the context of parameters production and operation composition. Additionally, this is the first time, to our knowledge, an analysis is conducted on the topology of semantic networks, and consequently on the comparison with syntactic networks.

The study presented in this paper can be improved according to two directions. First, the collection we used is based on a set of real-world WS descriptions, but part of them was generated through resampling, so it cannot be considered as perfectly realistic. As a matter of fact, no other publicly available collection provides both syntactic and semantic descriptions for the same services (Cherifi, 2009), which is an indispensable prerequisite to a consistent comparison. The only solution we can see is to constitute our own collection, by semantically annotating a set of real syntactic descriptions. Second, we used a selected set of matching functions to extract the dependency networks. Many other functions exist, in particular more flexible syntactic distances (Cohen et al., 2003) can be used to perform less strict comparisons of the parameters names. This could have significant implications on the resulting network properties, since it is directly related to the amount of false positives (nodes irrelevantly connected) and false negatives (nodes irrelevantly disconnected).

Besides those improvements on data and matching functions, we plan to extend our work in two ways. First, we want to analyze in greater details the partial overlapping observed between communities and domains. A related point is to test whether properties observed for the whole network are also valid for domains or sets of domains. Second, to compare the observations we made in this work on the parameters networks and in a parallel work on operations networks (Cherifi et al., 2010), we will extract and study networks at the service granularity level.